\documentstyle[12pt]{article}

\advance \topmargin by -\headheight
\advance \topmargin by -\headsep
     
\evensidemargin \oddsidemargin
\def\nonu{\nonumber}

\def\lab#1{\label{eq:#1}} 
\def\br{\begin{eqnarray}}
\def\er{\end{eqnarray}}
\def\be{\begin{equation}}
\def\ee{\end{equation}}
\def\0{\nonumber}
\def\lb{\lbrack}
\def\rb{\rbrack}
\def\({\left(}
\def\){\right)}

\relax

\def\a{\alpha}
\def\b{\beta}

\def\d{\delta} 
\def\D{\Delta}
\def\eps{\epsilon}
\def\g{k}

\def\l{\lambda}

\def\o{\over}

\def\pa{\partial}
\def\pr{\prime}

\def\s{\sigma}

\def\lie{{\cal G}}

\def\rlx{\relax\leavevmode}
\def\inbar{\vrule height1.5ex width.4pt depth0pt}
\def\IZ{\rlx\hbox{\sf Z\kern-.4em Z}}
\def\IR{\rlx\hbox{\rm I\kern-.18em R}}
\def\IC{\rlx\hbox{\,$\inbar\kern-.3em{\rm C}$}}
\def\one{\hbox{{1}\kern-.25em\hbox{l}}}

\def\NPB#1#2#3{{\sl Nucl. Phys.} {\bf B#1} (#2) #3}

\def\PRD#1#2#3{{\sl Phys. Rev.} {\bf D#1} (#2) #3}
\def\PRA#1#2#3{{\sl Phys. Rev.} {\bf A#1} (#2) #3}
\def\PLA#1#2#3{{\sl Phys. Lett.} {\bf #1A} (#2) #3}

\def\JMP#1#2#3{{\sl J. Math. Phys.} {\bf #1} (#2) #3}

\def\JPA#1#2#3{{\sl J. Physics} {\bf A#1} (#2) #3}

\def\JPSJ#1#2#3{{\sl J. of Phys. Soc. of Japan} {\bf #1} (#2) #3}
\def\JGP#1#2#3{{\sl J. of Geom. and Phys. } {\bf #1} (#2) #3}

\date{}    
\begin{document}


\begin{center}
{\large\bf    A Class of Mixed Integrable Models}
\end{center}
\normalsize
\vskip .4in

\begin{center}
J.F. Gomes$^1$, G.R. de  Melo$^{1,2}$ and A.H. Zimerman$^1$ 

\par \vskip .1in \noindent
$^1$ Instituto de F\'{\i}sica Te\'{o}rica-UNESP \\
Rua Pamplona 145\\
fax (55)11 31779080\\
01405-900 S\~{a}o Paulo, Brazil
\par \vskip .3in

\par \vskip .1in \noindent
$^2$ Faculdade Metropolitana de Cama\c{c}ari - FAMEC,\\
Av. Eixo Urbano Central, Centro,\\
 42800-000, Cama\c{c}ari, BA,
Brazil.
\par \vskip .3in

\end{center}
\baselineskip=1cm
\begin{abstract}
The algebraic structure of the  integrable mixed mKdV/sinh-Gordon model   is discussed and \textit{}extended to the AKNS/Lund-Regge model and to its corresponding supersymmetric versions.  The integrability of the models is guaranteed from the zero curvature representation and some soliton solutions are discussed.
\end{abstract}

\section{Introduction}

The mKdV and the sine-Gordon equations 
are   non-linear differential equations  belonging to    the same integrable 
hierarchy representing different time evolutions \cite{chodos}.  
The structure of its soliton solutions present the same   functional form in terms  of 
\br
\rho = e^{kx + k^nt_n}, 
\label{space}
\er
which carries the space-time dependence.  
  Solutions of different equations within 
the same hierarchy  differ  only by the factor $k^nt_n$ in $\rho$. 
For instance $n=3$ correspond to the mKdV equation and $n=-1$ to the sinh-Gordon.
For $n>0$  a systematic construction of integrable hierarchies  can be solved and classified  according to
 a  decomposition of an affine Lie algebra, $\hat \lie$ and a choice of a semi-simple constant element $E$ 
 (see \cite{nissimov} for review).  Such framework was  shown to be derived from the 
 Riemann-Hilbert decomposition which later, was shown to incorporate  negative grade  isospectral flows $n<0$ \cite{modugno} as well.

The mixed system 
\br
 \phi_{xt} = {{\a_3}\o {4}} \( \phi_{xxxx} - 6 \phi^2_x \phi_{xx}\) +2 \eta \sinh (2\phi) 
\label{s1a}
\er
is a non-linear differential equation  which represents the well 
known mKdV equation  for $\eta =0$ ( $v= -\pa_x \phi$)  and  the sinh-Gordon equation  
for $\a_3=0$. It  was introduced   in \cite{konno} where, employing   the inverse scattering method,
  multi soliton solutions were constructed by modification of time dependence in $\rho$.
  Solutions (multi soliton) were also considered in \cite{chen} by Hirota's method. Moreover,  
 two-breather solution was discussed in \cite{leblond} in connection with  few-optical-cycle pulses in transparent media.
The soliton solutions obtained in  
\cite{konno}, \cite{chen} and \cite{leblond} indicates integrability of the mixed model (\ref{s1a}).

In this paper we consider the mixed system  mKdV/sinh-Gordon (\ref{s1a}) within the  zero curvature representation.  
We show that a systematic solution  for the mixed model  is obtained by the dressing method and  
a specific choice of vacuum solution.  Such  formalism is extended to the mixed AKNS/Lund-Regge and to its supersymmetric  versions as well.

In the last section we discuss the coupling of higher positive and negative flows generalizing the examples  given previously.

\section{ The  mixed mKdV/sinh-Gordon model}
Let us consider a non-linear system composed of a mixed sinh-Gordon and mKdV equation  given by eqn. (\ref{s1a}) and 
the following zero curvature representation,
\br
\lb \pa_x + E^{(1)} + A_0, \pa_t + D_3^{(3)} + D_3^{(2)} +D_3^{(1)} +D_3^{(0)} +D_3^{(-1)} \rb = 0
\label{s2}
\er 
where $E^{(2n+1)} = \l ^n (E_{\a} + \l E_{-\a}), \quad A_0 = v h$ and $E_{\pm \a}$ and $ h$ are $sl(2)$ generators satisfying $\lb h, E_{\pm \a} \rb = \pm 2E_{\pm \a}, \quad \lb E_{\a}, E_{-\a} \rb= h$.  According to the grading operator $Q= 2\l {{d}\o {d\l}} + {1\o 2} h$,   $D_3^{(j)}$  is a  graded  $j$ Lie algebra valued and eqn. (\ref{s2}) decomposes into 6 independent equations (decomposing grade by grade):
\br
\lb E, D_3^{(3)}\rb &=& 0, \nonu \\
\lb E, D_3^{(2)}\rb  + \lb A_0, D_3^{(3)} \rb  +\pa_x D_3^{(3)} &=& 0, \nonu \\
\lb E, D_3^{(1)}\rb  + \lb A_0, D_3^{(2)}  \rb + \pa_x D_3^{(2)} &=& 0, \nonu \\
\lb E, D_3^{(0)}\rb  + \lb A_0, D_3^{(1)}  \rb + \pa_x D_3^{(1)} &=& 0, \nonu \\
\lb E, D_3^{(-1)}\rb  + \lb A_0, D_3^{(0)} \rb + \pa_x D_3^{(0)} - \pa_t A_0 &=& 0, \nonu \\
 \lb A_0, D_3^{(-1)}\rb + \pa_x D_3^{(-1)} &=& 0.
 \label{s3}
 \er
 where $E \equiv E^{(1)}$.
In order to solve (\ref{s3}) let us propose
\br
D_3^{(3)}&=& \a_3 \( \l E_{\a} + \l^2 E_{-\a} \) + \b_3 \( \l E_{\a} - \l^2 E_{-\a} \), \nonu \\
D_3^{(2)}&=& \s_2 \l h, \nonu \\
D_3^{(1)}&=& \a_1 \(  E_{\a} + \l E_{-\a} \) + \b_1 \(  E_{\a} - \l E_{-\a} \), \nonu \\
D_3^{(0)}&=& \s_0 h. 
\label{s4}
\er
Substituting (\ref{s4}) in (\ref{s3})  we obtain $\b_3 = 0, \quad \a_3 = const$ and 
\br
 \b_1 = {{\a_3}\o 2} v_x, \qquad \a_1 = -{{\a_3}\o 2} v^2, \qquad \s_0 = {{\a_3}\o {4}}(v_{xx} -2v^3), \qquad 
\s_2 =\a_3v. 
\label{s5}
\er
In order to solve the last eqn. in (\ref{s3}) we parametrize
\br
A_0 = -\pa_x B B^{-1} = -\pa_x \phi h, \qquad B = e^{\phi h}
\label{a0}
\er
and 
\br
D_3^{(-1)} = \eta B E^{(-1)} B^{-1} = \eta \l^{-1} (e^{2\phi} E_{\a} + \l e^{-2\phi} E_{-\a})
\label{d-1}
\er
The zero grade projection in (\ref{s3}) yields the  time evolution equation (\ref{s1a}).  
Notice that in order to solve  the last eqn. (\ref{s4}) we have introduced the sinh-Gordon variable $\phi$ in (\ref{a0}) and in (\ref{d-1}) such that $v = -\pa_x \phi$.

 Let us now recall some basic aspects of the dressing  method which provides systematic construction  
 of soliton solutions.   The zero curvature representation implies in a pure gauge configuration.  
 In particular,  the vacuum is obtained by setting $\phi_{vac}  = 0$ or $v_{vac} = 0$ which, when  in (\ref{s2}) 
 implies,
 \br
 \pa_x T_0 T_0^{-1} = E^{(1)}, \qquad  \pa_t T_0 T_0^{-1} = \a_3E^{(3)} + \eta E^{(-1)}
 \label{s6}
 \er
 and after integration 
 \br
 T_0 = \exp \( t(\a_3E^{(3)} + \eta E^{(-1)})\) \exp \(xE^{(1)}\), \qquad E^{(2n+1)} = \l^n(E_{\a} + \l E_{-\a}).
 \label{s7}
 \er
 If we identify
 $v= -\pa_x \phi$ 
 eqn. (\ref{s1a}) 
 represents  a coupling of mKdV and sinh-Gordon equations and becomes a pure  
 mKdV  when $\eta =0$ and pure sinh-Gordon  when $\a_3 =0$.  Tracing back those two limits from (\ref{s5}) and (\ref{d-1})
 it becomes clear that 
 the sinh-Gordon limit ($\eta =0$) in (\ref{s1a}) is responsible for the vanishing of $D_3^{(-1)}$.
 On the other hand,  $\a_3=0$ implies      $D_3^{(j)} =0, j=0, \cdots 3$.
 Inspired by the dressing method  for constructing soliton solutions of integrable 
hierarchies (see for instance \cite{laf}) and the fact that the $n-$th member of the
hierachy is associated to the time evolution parameter $k_i^n t_n$ ($n=3$ for mKdV 
and $n=-1$ for sinh-Gordon) it is natural to propose soliton solutions  based on the
modified space-time dependence
\br
 \rho_i = \exp \(2k_i x +2(\a_3k_i^3 +{\eta}/ {k_i})t\).
 \label{dep}
 \er
It therefore follows that 
the general structure of the 1-,  2- and 3-soliton solutions  is
respectively given by (after $\phi \rightarrow i\phi$)
\br
\phi_{1-sol} &=& iln \( {{1- a_1 \rho_1 }\o {1+a_1 \rho_1}}\), \nonu \\
\phi_{2-sol} &=& iln \( {{1- a_1 \rho_1 -a_2 \rho_2 +a_1a_2 a_{12} \rho_1\rho_2}\o {1+a_1 \rho_1 +a_2 \rho_2 +a_1a_2a_{12} \rho_1\rho_2}}\), \nonu \\
\phi_{3-sol} &=& iln \({{1- \sum_{i=1}^3 a_i \rho_i + \sum_{i<j=1}^{3} a_ia_j a_{ij} \rho_i\rho_j - a_1a_2a_3a_{12}a_{13}a_{23}\rho_1\rho_2\rho_3}\o {1+ \sum_{i=1}^3 a_i \rho_i + \sum_{i<j=1}^{3} a_ia_j a_{ij} \rho_i\rho_j + a_1a_2a_3a_{12}a_{13}a_{23}\rho_1\rho_2\rho_3}}\)
\label{s6}
\er
where $ a_1, a_2$ are  constants and $a_{ij} = ({{k_i-k_j}\o {k_i+k_j}})^2$. 

More general solutions  (N-solitons and breathers ) were found in refs. \cite{konno}, \cite{chen} and \cite{leblond}  with same time dependence as in (\ref{dep}).

\section{The Mixed AKNS/Lund-Regge Model}

Let us   consider another example   involving $\lie = sl(2)$ and  homogeneous gradation 
$Q= \l {{d}\o {d\l}}, \quad E^{(n)} = \l^n h, \;\; E= E^{(1)}$ and $A_0 = qE_{\a} + rE_{-\a}$ and 
 the zero curvature representation of the form
\br
\lb \pa_x + E + A_0, \pa_t  + D_2^{(2)} +D_2^{(1)} +D_2^{(0)} +D_2^{(-1)} \rb = 0
\label{s8}
\er 
According to gradation $Q$, propose
\br
 D_2^{(j)} &=& \l^j \( \a_j E_{\a} +\b_j E_{-\a} + \s_j  h\), \qquad j=-1,0,1,2
 \label{s9}
 \er
 In order to find solution for (\ref{s8}) we introduce  variables $\tilde \psi$ and $\tilde \chi$  \cite{jpal},
 \br
  A_0 = qE_{\a} + rE_{-\a}= -\pa _xB B^{-1}, \qquad D_2^{(-1)} = \eta BE^{(-1)} B^{-1}, \qquad 
  B =  e^{\tilde \chi E_{-\a} } e^{\phi h } e^{\tilde \psi E_{\a} }
 \label{s13}
 \er
 which defines
 \br
 q=-\pa_x \tilde \psi e^{2\phi}, \qquad r =\tilde \chi^2 \pa_x \tilde \psi e^{2\phi} -\pa_x \tilde \chi
 \label{akns}
 \er 
 together with the subsidiary conditions for the non-local auxiliary field $\phi$,
 \br
 Tr \( \pa_xBB^{-1}h\) = \pa_x \phi  - \tilde \chi \pa_x \tilde \psi e^{2\phi}  =0, \qquad 
 Tr \( B^{-1}\pa_tBh\) = \pa_t \phi  - \tilde \psi \pa_t \tilde \chi e^{2\phi} =0.
 \label{s14}
 \er
 Solution of constraints (\ref{s14}) leads to natural variables  \cite{wigner}
 \br 
  \psi = \tilde \psi e^{\phi}, \qquad  \chi = \tilde \chi e^{\phi}.
 \label{variables}
 \er
 Inserting (\ref{s9}) into (\ref{s8}) and collecting powers of $\l$,  we find solution in terms of non-local fields $ \psi$ and $ \chi$ 
\br 
\s_2= const, \qquad  \b_2 = \a_2 = 0, \qquad  \s_1 = 0,\qquad \s_0 = -1/2 \s_2 r q \nonu \\
\b_1 = \s_2 r, \qquad \a_1 = \s_2q, \qquad \a_0 = -1/2 \s_2q_x, \qquad \b_0 = 1/2 \s_2 r_x, \nonu \\
\a_{-1} = -2 \eta  \psi e^{\phi}, \qquad \b_{-1} = 2\eta ( \chi + \psi  \chi^2)e^{-\phi}, \qquad \s_{-1} = \eta (1+2  \psi  \chi) 
\label{s10}
\er
leading to the equations of motion
\br
q_t +{1\o 2} \s_2\(q_{xx} -2q^2 r\) -2\a_{-1}=0, \nonu \\
r_t -{1\o 2} \s_2\(r_{xx} -2r^2 q\) +2\b_{-1}=0,
\label{s11}
\er
where $q$ and $r$ in variables $ \psi$ and $\chi$ reads
\br
q=- {{\pa_x  \psi}\o {1+ \psi  \chi}}e^{\phi} \qquad r = - \pa_x  \chi e^{-\phi}.
\label{rq}
\er
Equations (\ref{s11}) represent  a mixed system of AKNS 
 (for $\eta =0$, $\a_{-1} = \b_{-1} =0$) in variables $q,r$   and the relativistic Lund-Regge (for $\s_2=0$) in variables $ \psi,  \chi$.
 \br
  \pa_t \({{\pa_x  \psi}\o {\D}}\) +  \psi {{\pa_t  \chi \pa_x  \psi}\o {\D^2}} + 4 \eta  \psi &=&0, \nonu \\
  \pa_x \({{\pa_t  \chi}\o {\D}}\) +  \chi {{\pa_t  \chi \pa_x  \psi}\o {\D^2}} + 4 \eta  \chi &=&0. 
  \label{lr}
  \er
 Again the terms proportional to $\a_{-1} $ and $\b_{-1}$  originate from the  contribution of 
  $ D^{(-1)}_2 =\eta B E^{(-1)} B^{-1}$ in (\ref{s8}) and the vacuum configuration  is obtained  
  for $\psi_{vac} = \chi_{vac} = q_{vac}=r_{vac}=0$.  The model is now characterized by $E^{(n)} = \l^n h$ 
 and the vacuum solution of (\ref{s8}) yield
 \br
 T_0 = \exp \( t(\s_2E^{(2)} + \eta E^{(-1)})\) \exp \(xE^{(1)}\).
 \label{s15}
 \er
 and therefore the space-time  dependence in $\rho_i$ comes in the form 
 \br \rho_i = \exp \(2k_i x +2(\s_2k_i^2 +{\eta}/ {k_i})t\).
 \label{s16}
 \er
 We have checked the solution for the composite model (\ref{s11}) to agree with the functional form of the one proposed in ref. \cite{wigner}  with modified space-time dependence given by (\ref{s16}), i.e.,
 \br
  \psi = {{b \rho_2^{-1}}\o {1+{{k_1}\o {k_2}}\Gamma \rho_1\rho_2^{-1}}}, 
 \qquad  
  \chi = {{a \rho_1\o {1+{{k_1}\o {k_2}}\Gamma \rho_1\rho_2^{-1} }}}, \qquad 
 e^{-\phi} = {{1+{{\g_1} \o {\g_2}}\Gamma \rho_1\rho_2^{-1}\o 
 {1+\Gamma \rho_1\rho_2^{-1}}}} \nonu \\
 \label{sol1}
\er 
where $a$ and $b$ are  constants, $\Gamma = {{ab  \g_2^2}\o {(\g_1 -\g_2 )^2}}$.
  In terms of AKNS field variables, from (\ref{rq}) we find
\br
r = - {{2a \g_1 \rho_1}\o { 1+ {{ab \g_1 \g_2 }\o {(\g_1 -\g_2 )^2}}\rho_1\rho_2^{-1}}}, \quad 
q =  {{2b \g_2 \rho_2^{-1} }\o { 1+ {{ab \g_1 \g_2 }\o {(\g_1 -\g_2 )^2}}\rho_1\rho_2^{-1}}}. 
\label{sol2}
\er

 \section{The Supersymmetric mKdV/Sinh-Gordon Model}

 Following the same line of reasoning, we now consider algebraic structures with half integer gradation \cite{np}.
 Let $\lie = sl(2,1)$, $Q = 2\l{{d}\o {d\l}} + {1\o 2}h$ and 
 $E^{(1) }=\l^{1/2}(h_1 + 2h_2) -( E_{\alpha _{1}}+\l E_{-\alpha _{1}}) $.
 The graded structure can be decomposed as follows (see appendix of ref. \cite{pla}) for instance),
 \br
 {\cal K}_{Bose} = \{ K_1^{(2n+1)} &=& -(E_{\a_1}^{(n)} +E_{-\a_1}^{(n+1)}), \quad K_2^{(2n+1)} = \mu_2\cdot H^{(n+1/2)}\}, \nonu \\
{\cal M}_{Bose} = \{ M_1^{(2n+1)} &=& -E_{\a_1}^{(n)} + E_{-\a_1}^{(n+1)}, \qquad M_2^{(2n)} = h_1^{(n)} = \a_1 \cdot H^{(n)}\},\nonu \\
 {\cal K}_{Fermi} = \{ F_1^{(2n+3/2)} &=& (E_{\a_1+\a_2}^{(n+1/2)} -E_{\a_2}^{(n+1)} )+(E_{-\a_1-\a_2}^{(n+1)} - E_{-\a_2}^{(n+1/2)} ), \nonu \\
F_2^{(2n+1/2)} &=& -(E_{\a_1+\a_2}^{(n)} -E_{\a_2}^{(n+1/2)} )+(E_{-\a_1-\a_2}^{(n+1/2)} - E_{-\a_2}^{(n)} )\}, \nonu \\
{\cal M}_{Fermi} = \{ G_1^{(2n+1/2)} &=& ( E_{\a_1+\a_2}^{(n)} + E_{\a_2}^{(n+1/2)})+( E_{-\a_1-\a_2}^{(n+1/2)}+ E_{-\a_2}^{(n)}), \nonu \\
G_2^{(2n+3/2)} &=& -(E_{\a_1+\a_2}^{(n+1/2)}   + E_{\a_2}^{(n+1)} )+( E_{-\a_1-\a_2}^{(n+1 )} +E_{-\a_2}^{(n+1/2)}  )\}, \label{s17}
 \er
where we have denoted $E_{\pm \a}^{(n)} = \l^n E_{\pm \a}$ and $H^{(n)} = \l^n H$ and $\a_i, \mu_i, \; i=1,2 $ are respectively the simple roots and fundamental weights of $sl(2,1)$. In (\ref{s17}) we have denoted ${\cal K} = {\cal K}_{bose} \cup {\cal K}_{Fermi}$ to be the Kernel of $E^{(1)}$, i.e., $[ E^{(1)}, {\cal K}] =0$ and ${\cal M}$ is its complement.
  The Lax  operator is constructed as
 \br
 L = \pa_x + E^{(1)} + A_{1/2}+ A_0, \qquad A_0 = v M_2^{(0)}, \qquad A_{1/2} = \bar \psi G_1^{(1/2)}.
 \label{s18}
 \er
and the zero curvature representation  reads
 \br
\lb \pa_x + E^{(1)} + A_{1/2} + A_0, \pa_t + D_3^{(3)} + D_3^{(5/2)} +\cdots +D_3^{(-1/2)} +D_3^{(-1)} \rb = 0
\label{s19}
\er 
In order to solve   for the lowest grades $-1, -1/2$ of eqn. (\ref{s19}) we introduce  the parametrization
\br
D_3^{(-1)} = \eta B E^{(-1)} B^{-1}, \qquad A_0 = -\pa_x BB^{-1}, \qquad B= e^{\phi M_2^{(0)}}
\label{t1}
\er
together with the change of variables
\br
D_3^{(-1/2)} = B j_{-1/2} B^{-1}, \qquad j_{-1/2} = \psi G_2 ^{(-1/2)}
\label{t2}
\er
Propose solution of the form
\begin{eqnarray}
D_{3}^{(3)}&=& \a_3 \(h_1^{(3/2)} + 2h_2^{(3/2)} - E_{\alpha _{1}}^{(1)}- E_{-\alpha _{1}}^{(2)} \), \nonu \\
D_{3}^{(0)}&=&\a_{1}M_{2}^{(0)}, \nonumber \\
D_{3}^{(1/2)}&=& \b_{1}G_{1}^{(1/2)}+\b_{2}F_{2}^{(1/2)},\nonumber \\
D_{3}^{(1)}&=&\s_{1}M_{1}^{(1)}+\s_{2}K_{1}^{(1)}+\s_{3}K_{2}^{(1)}, \nonumber \\
D_{3}^{(3/2)}&=&\d_{1}G_{2}^{(3/2)}+\d_{2}F_{1}^{(3/2)},\nonumber\\
D_{3}^{(2)}&=&\mu_{1}M_{2}^{(2)}, \label{ds} \\
D_{3}^{(5/2)}&=&\nu_{1}G_{1}^{(5/2)}+\nu_{2}F_{2}^{(5/2)},\nonumber \\
D_{3}^{(-1/2)}&=& \b_{-1}G_{1}^{(-1/2)}+\b_{-2}F_{1}^{(-1/2)},\nonumber \\
D_{3}^{(-1)}&=&\s_{-1}M_{1}^{(-1)}+\s_{-2}K_{1}^{(-1)}+ \s_{-3} K_{2}^{(-1)}. \nonumber 
\end{eqnarray}
 where the coeffcients are given  by
 \br
 \a_1 &=& {1\o 4} \pa_x^2 v + {3\o 4} v\bar \psi \pa_x \bar \psi - {1\o 2} v^3, \qquad \b_1= {1\o 4} \pa_x^2 \bar \psi - {1\o 2} v^2 \bar \psi, \qquad \b_2 = {1\o 4}( v\pa_x \bar \psi - \bar \psi \pa_x v), \nonu \\
 \s_1 &=& {1\o 2} \pa_x v, \qquad \s_2 = {1\o 2} (\bar \psi \pa_x \bar \psi - v^2), \qquad \s_3 = -{1\o 2} \bar \psi \pa_x \psi\qquad \d_1 =-{1\o 2}\pa_x \bar \psi, \qquad \d_2 = -{1\o 2} v \bar \psi, \nonu \\
 \mu_1 &=& v, \qquad \nu_1 = \bar \psi, \qquad  \nu_2 =0, \qquad \b_{-1} = \psi \cosh{\phi}, \qquad \b_{-2} = -\psi \sinh \phi, \nonu \\
 \s_{-1} &=& \eta \sinh 2 \phi, \qquad \s_{-2}= \eta  \cosh 2\phi, \qquad \s_{-3} = \eta 
 \label{s20}
 \er
 where $\a_3$ and $\eta $ are arbitrary constants.  
 The equations of motion are given by grades $0, \pm 1/2$ projections of  (\ref{s19}), i.e.,
 \begin{eqnarray}
\partial_{t}\pa_x \phi &=&\frac{\a_{3}}{4}\left[\partial_{x}^{4}\phi -6(\pa_x \phi)^{2} \partial_{x}^2\phi +3\bar{\psi}\partial_{x}(\pa_x \phi \partial_{x}\bar{\psi})\right]+ 2\eta  \left[ \sinh(2\phi) + \bar \psi  \psi \sinh (\phi)\right], \nonu \\
\partial_{t_{3}}\bar{\psi}&=&\frac{\a_{3}}{4} \left[ \partial_{x}^{3}\bar{\psi}-3\pa_x \phi \partial_{x}(\pa_x\phi \bar{\psi})\right]+2 \eta  \psi \cosh(\phi),\nonu \\
\partial_{x} \psi &=&2  \bar{\psi} \cosh (\phi).
\label{s21}
\end{eqnarray}
Observe that for $\eta =0$ eqns. (\ref{s21})  corresponds to the $N=1$ super mKdV equation  if we identify $v=-\pa_x \phi$ and for $\a_3=0$ they correspond to the $N=1$ super sinh-Gordon.

The soliton solutions   are parametrized  in terms of tau functions as
 \br
 \phi = ln \( {{\tau_1}\o {\tau_0}}\), \qquad \bar \psi ={{ \tau_3}\o {\tau_1}} +{{\tau_2}\o {\tau_0}}
 \label{tau}
 \er
 The one-soliton solution for the $N=1$ super sinh-Gordon  and mKdV equations is given by
\begin{eqnarray}
\tau_{0}&=&1-\frac{1}{2}b_{1}{\rho}_{1},\qquad 
\tau_{1}=1+\frac{1}{2}b_{1}{\rho}_{1},\nonumber\\
\tau_{2}&=&c_{1}k_{2}{\rho}_{2}^{-1}+b_{1}c_{1}\sigma_{1,2}{\rho}_{1}{\rho}_{2}^{-1},\qquad
\tau_{3}=c_{1}k_{2}{\rho}_{2}^{-1}-b_{1}c_{1}\sigma_{1,2}{\rho}_{1}{\rho}_{2}^{-1},
\label{1-sol}
\end{eqnarray}
where  $\sigma_{1,2}={1\o 2} k_{2}\frac{(k_{1}+k_{2})}{(k_{1}-k_{2})}$, $b_1, c_1$ are bosonic and grassmaniann constants respectively 
 and  $\rho_i$ carries the space-time dependence for the sinh-Gordon and mKdV respectively, 
\br
\rho_i^{mKdV} = \exp \(2k_{i}x+2 ( \a_{3} k_i^{3}  ) t\), \qquad \rho_i^{s-G} = \exp \(2k_{i}x+2 ( {\eta \o { k_i}}  ) t\)
\label{rho-sg}
\er
Notice however that the introduction of the $D_{-1}^{(-1/2)}$ and $D_{-1}^{(-1)}$ 
terms  changes the vacuum configuration such that 
\br T_0 = \exp (x  E^{(1)}) \exp (\a_3 E^{(3)} + \eta E^{(-1)}) t)
\label{s22}
\er 
which induces   modification in the space-time dependence of eqns. (\ref{s21}) as
\br
\rho_i = \exp \(2k_{i}x \) \exp  \( 2(\a_{3} k_i^{3} + {{\eta}\o { k_i}} ) t\). 
\label{rho}
\er
 In fact we have verified explicitly that (\ref{1-sol}) with
  (\ref{rho}) satisfies the equations of motion (\ref{s21}). 
   The same was verified for the two soliton solution 
 \begin{eqnarray}
 \tau_{0} &=&1-\frac{1}{2}b_{1}{\rho}_{1}-\frac{1}{2}b_{2}{\rho}_{2}+b_{1}b_{2}{\rho}_{1}{\rho}_{2}\alpha_{1,2}\nonumber \\
&+&c_{1}c_{2}{\rho}_{3}^{-1}{\rho}_{4}^{-1}(\beta_{3,4}-b_{1}{\rho}_{1}\delta_{1,3,4}-b_{2}{\rho}_{2}\delta_{2,3,4}+b_{1}b_{2}{\rho}_{1}{\rho}_{2}\theta_{1,2,3,4}),\nonu \\
\tau_{1} &=&1+\frac{1}{2}b_{1}{\rho}_{1}+\frac{1}{2}b_{2}{\rho}_{2}+b_{1}b_{2}{\rho}_{1}{\rho}_{2}\alpha_{1,2}\nonumber\\
&+&c_{1}c_{2}{\rho}_{3}^{-1}{\rho}_{4}^{-1}(\beta_{3,4}+b_{1}{\rho}_{1}\delta_{1,3,4}+b_{2}{\rho}_{2}\delta_{2,3,4}+b_{1}b_{2}{\rho}_{1}{\rho}_{2}\theta_{1,2,3,4}),\nonumber \\
\tau_{2}
&=&c_{1}{\rho}_{3}^{-1}\left(k_{3}+b_{1}{\rho}_{1}\sigma_{1,3}+b_{2}{\rho}_{2}\sigma_{2,3}+b_{1}b_{2}{\rho}_{1}{\rho}_{2}\lambda_{1,2,3}\right)\nonumber \\
&+&c_{2}{\rho}_{4}^{-1}\left(k_{4}+b_{1}{\rho}_{1}\sigma_{1,4}+b_{2}{\rho}_{2}\sigma_{2,4}+b_{1}b_{2}{\rho}_{1}{\rho}_{2}\lambda_{1,2,4}\right),\nonumber \\
\tau_{3}
&=&c_{1}{\rho}_{3}^{-1}\left(k_{3}-b_{1}{\rho}_{1}\sigma_{1,3}-b_{2}{\rho}_{2}\sigma_{2,3}+b_{1}b_{2}{\rho}_{1}{\rho}_{2}\lambda_{1,2,3}\right)\nonumber \\
&+&c_{2}{\rho}_{4}^{-1}\left(k_{4}-b_{1}{\rho}_{1}\sigma_{1,4}-b_{2}{\rho}_{2}\sigma_{2,4}+b_{1}b_{2}{\rho}_{1}{\rho}_{2}\lambda_{1,2,4}\right),
\end{eqnarray}
where
\begin{eqnarray}
\alpha_{1,2}&=&\frac{1}{4}\frac{(k_{1}-k_{2})^{2}}{(k_{1}+k_{2})^{2}},\nonumber \\
\beta_{3,4}&=&k_{3}k_{4}\frac{(k_{3}-k_{4})}{(k_{3}+k_{4})^{2}},\nonumber \\
\delta_{j,3,4}&=&\frac{k_{3}k_{4}}{2}\frac{(k_{3}-k_{4})}{(k_{3}+k_{4})^{2}}\frac{(k_{j}+k_{3})}{(k_{j}-k_{3})}\frac{(k_{j}+k_{4})}{(k_{j}-k_{4})} \qquad (j=1,2),\nonumber \\
\sigma_{j,k}&=&\frac{k_{k}}{2}\frac{(k_{j}+k_{k})}{(k_{j}-k_{k})} \qquad (j=1,2) \qquad (k=3,4),\nonumber\\[12pt]
\lambda_{1,2,j}&=&\frac{k_{j}}{4}\frac{(k_{1}-k_{2})^{2}}{(k_{1}+k_{2})^{2}}\frac{(k_{1}+k_{j})}{(k_{1}-k_{j})}\frac{(k_{2}+k_{j})}{(k_{2}-k_{j})}, \qquad (j=3,4),\nonumber\\[12pt]
\theta_{1,2,3,4}&=&\frac{k_{3}k_{4}}{4}\frac{(k_{1}-k_{2})^{2}}{(k_{1}+k_{2})^{2}}\frac{(k_{1}+k_{3})}{(k_{1}-k_{3})}\frac{(k_{2}+k_{3})}{(k_{2}-k_{3})}\nonumber\frac{(k_{3}-k_{4})}{(k_{3}+k_{4})^{2}}\frac{(k_{1}+k_{4})}{(k_{1}-k_{4})}\frac{(k_{2}+k_{4})}{(k_{2}-k_{4})},\nonumber\\
\end{eqnarray}
 $b_1, b_2$ are bosonic constants and $c_1, c_2$ are Grassmaniann constants with $\rho_i $ given by (\ref{rho}).

 \section{The Supersymmetric Lund-Regge/AKNS model}
 
 In this section we consider the Lie superalgebra $\lie = sl(2,1)$ with homogeneous gradation, $Q= \l {{d}\o {d\l}}$ and (see for instance \cite{silka})
 \br
 E^{(n)} = (\a_1 + \a_2 )\cdot H^{(n)}, \qquad \a_1, \a_2 \rm{\quad are \;\;  simple \;\;  roots \;\;  of \;\; sl(2,1)}.
 \label{e}
 \er 
 The Lax operator is then
 \br
 L
 = \pa_x + E^{(1)} + A_0, \qquad A_0= b_1E_{\a_1}+  \bar b_1E_{-\a_1}+F_1 E_{\a_2} + \bar {F}_1 E_{-\a_2}
 \label{e2}
 \er
 We search for solution of 
 \br
\lb \pa_x + E^{(1)}  + A_0, \pa_t + D_2^{(2)} + D_2^{(1)} +D_2^{(0)} +D_2^{(-1)} \rb = 0
\label{e3}
\er 
   Decomposing (\ref{e3}) grade by grade, we find 
 \br
 D_2^{(2)} &=&  a_2\l^2  \a_1 \cdot H,\nonu \\
 D_2^{(1)} &=&  g_1 \l E_{\a_1} + m_1 \l E_{-\a_1} + n_1 \l E_{-\a_2} + o_1\l  E_{\a_2}\nonu \\
 D^{(0)}_{{2 \cal M}} &=& g_0 E_{\a_1} + m_0 E_{-\a_1} + n_0 E_{-\a_2}+ o_0 E_{\a_2}   ,\nonu \\
 D^{(0)} _{{2 \cal K}}&=& a_0 \a_1 \cdot H + c_0\a_2 \cdot H + d_0 E_{\a_1+\a_2} + e_0 E_{-\a_1-\a_2} .
\label{e4}
\er
where $D_2^{(0)} = D^{(0)}_{{2 \cal M}}+D^{(0)}_{{2 \cal K}}$ and 
\br
g_1 = a_2 b_1 \qquad m_1 = a_2 \bar b_1, \qquad o_1 = a_2 F_1, \qquad n_1 = a_2 \bar F_1, \nonu \\
g_0 = a_2 \pa_x b_1, \qquad m_0 = -a_2 \pa_x \bar b_1, \qquad n_0 = a_2 \pa_x \bar F_1, \qquad o_0 = -a_2 \pa_x F_1, \nonu \\
d_0 = -a_2 F_1 b_1 , \qquad   e_0 = -a_2 \bar F_1 \bar b_1, \qquad a_0 = -a_2 b_1 \bar b_1, \qquad c_0 = -a_2 (b_1 \bar b_1 +F_1 \bar F_1 )
\nonu 
\er
In order to solve the grade $-1$  projection of eqn. (\ref{e3})
we introduce the $sl(2,1)$ variables \cite{silka} as
\br
A_0= -\pa_x BB^{-1} =  b_1E_{\a_1}+  \bar b_1E_{-\a_1}+F_1 E_{\a_2} + \bar {F}_1 E_{-\a_2}, 
\label{f1}
\er 
where 
\br
B = e^{\tilde \chi E_{-\a_1}}e^{\tilde { f_1} E_{-\a_1-\a_2}}e^{\tilde { f_2} E_{\a_2}}e^{\varphi_1 (\a_1+\a_2) \cdot H 
 - \varphi_2 \a_2 \cdot H}
 e^{\tilde {g_2} E_{-\a_2 }}e^{\tilde {g_1} E_{\a_1+\a_2}}e^{\tilde \psi E_{\a_1}}
 \lab{d1}
 \er
 and 
 \br
  D^{(-1)}_{2\; \cal M} = \eta B E^{(-1)}B^{-1} &=& -\eta \psi e^{{1\o 2} (\phi_1 + \phi_2)}\l^{-1} E_{\a_1} + \eta f_2 (1+ \psi \chi) e^{-{1\o 2} \phi_1}\l^{-1} E_{\a_2}, \nonu \\
  &+& \eta (\chi +  f_1  f_2 +\psi \chi^2 + \psi \chi  f_1  f_2)e^{-{1\o 2} (\phi_1 + \phi_2)}\l^{-1}  E_{-\a_1} \nonu \\
  &-& \eta (g_2 + \psi  f_1 ) e^{{1\o 2} \phi_1}\l^{-1} E_{-\a_2}
  \label{d12}
  \er
written in the natural variables
\br
 \tilde \psi = \psi e^{-\frac{\varphi_{1}+\varphi_{2}}{2}}, \quad 
\tilde {g_{1}} =  g_{1}e^{-\frac{\varphi_{2}}{2}}, \quad
\tilde {f_{1}} =  f_{1}e^{-\frac{\varphi_{2}}{2}}\nonu \\
\tilde \chi =  \chi e^{-\frac{\varphi_{1}+\varphi_{2}}{2}}, \quad 
\tilde {g_{2}} =  g_{2}e^{-\frac{\varphi_{1}}{2}}, \quad
\tilde {f_{2}} =  f_{2}e^{-\frac{\varphi_{1}}{2}}.
\lab{d2}
 \er
 Here, $\psi, \chi, \varphi_i, i=1,2$ and $f_i, g_i, i=1,2$ are bosonic and fermionic fields respectively.
 The absence of Cartan subalgebra $h_1, h_2$ and $E_{\pm (\a_1+ \a_2)}$ (i.e. in ${\cal K}$) in the r.h.s. of (\ref{f1}) leads to the following subsidiary constraints
 \br
\pa_t {f}_{1}&=&\frac{1}{2}{f}_{1}\pa_t\varphi_{2}
+g_{2}[\pa_t\chi -\frac{1}{2}\chi (\pa_t\varphi_{1} +\pa_t\varphi_{2})], \nonu \\
\pa_t g_{1}&=&\psi\pa_t {f}_{2} +\frac{1}{2}g_{1} \pa_t\varphi_{2}
-\frac{1}{2}\psi {f}_{2}\pa_t\varphi_{1}, \nonu \\
 {\partial_x} {f}_{1}&=&\chi{\partial_x} g_{2}
+\frac{1}{2}{f}_{1} {\partial_x}\varphi_{2} -\frac{1}{2}\chi
g_{2}{\partial_x}\varphi_{1},\nonu \\
 {\partial_x} g_{1}&=&\frac{1}{2}g_{1}{\partial_x}\varphi_{2}
+{f}_{2}[{\partial_x}\psi -\frac{1}{2}\psi
({\partial_x}\varphi_{1} +{\partial_x}\varphi_{2})], \nonu \\
\pa_t\varphi_{1}&=&\frac{\psi [\pa_t\chi (1 +g_{2}{f}_{2})
+\frac{1}{2}\chi g_{2}\pa_t {f}_{2}]}{1 +\psi\chi (1
+\frac{5}{4}g_{2}{f}_{2})}, \nonu \\
\pa_t\varphi_{2}&=&\frac{\psi\pa_t\chi (1 +\frac{3}{2}g_{2}{f}_{2})
-g_{2}\pa_t {f}_{2} -\frac{1}{2}\psi\chi g_{2}\pa_t
{f}_{2}}{1 +\psi\chi (1 +\frac{5}{4}g_{2}{f}_{2})}, \nonu \\
 {\partial_x}\varphi_{1}&=&\frac{\chi
[{\partial_x}\psi (1 +g_{2}{f}_{2})
+\frac{1}{2}\psi{\partial_x}g_{2}{f}_{2}]}{1 +\psi\chi (1
+\frac{5}{4}g_{2}{f}_{2})}, \nonu \\
 {\partial_x}\varphi_{2}&=&\frac{\chi{\partial_x}\psi
(1 +\frac{3}{2}g_{2}{f}_{2}) +(\frac{1}{2}\psi\chi
+1){f}_{2}{\partial_x}g_{2}}{1 +\psi\chi (1
+\frac{5}{4}g_{2}{f}_{2})}.
\lab{nonlocal}
\er
 Moreover eqn. (\ref{f1}) yields
\br
\bar b_1  &=& \bar J_{-\a_1} = -\frac{e^{\frac{1}{2}(\varphi_1 +\varphi_2)}}
{1+f_2g_2}
  \(  \pa_x \psi - \frac{1}{2}\psi ( \pa_x \varphi_1 +  \pa_x \varphi_2 )\), \nonu \\
F_1   &=& \bar J_{-\a_2} = -e^{-\frac{1}{2}\varphi_1}\(  \pa_x f_2 + \frac{1}{2} f_2  \pa_x \varphi_1 \), \nonu \\
b_1  & = & -e^{-\frac{1}{2}(\varphi_1 +\varphi_2)}\(  \pa_x \chi +
\frac{1}{2}\chi ( \pa_x \varphi_1 + \pa_x \varphi_2)-\chi f_2  \pa_x g_2 -
\frac{1}{2}\chi   \pa_x \varphi_1 g_2 f_2 - e^{\frac{1}{2}\varphi_1}f_1 \bar J_{-\a_2} \right. \nonu \\
&+& \left. \chi^2 e^{-\frac{1}{2}(\varphi_1 -\varphi_2)}\bar J_{-\a_1}\),\nonu \\
\bar F_1  &=& -e^{\frac{1}{2}\varphi_1}\(  \pa_x g_2 -\frac{1}{2}g_2  \pa_x\textit{} \varphi_1 +
 e^{-\frac{1}{2}(\varphi_1 -\varphi_2)}f_1 \bar J_{-\a_1} \)
\label{rem}
\er
 Solving the zero grade component of (\ref{e3}), we find the equations of motion,
 \br
 & \pa_{t_2}b_1 + a_2\(\pa^2_{x} b_1 -2 \( b_1\bar b_1 +F_1\bar {F}_1  \) b_1\) + m_{-1}=0,  \nonu \\
& \pa_{t_2}\bar b_1 - a_2\(\pa^2_{x} \bar b_1 -2   \( b_1\bar b_1 +F_1\bar {F}_1 \)\bar  b_1\) - g_{-1}=0, \label{13a}\nonu  \\
& \pa_{t_2}F_1 - a_2(\pa^2_{x} F_1 -2 b_1 \bar b_1 F_1) - n_{-1} =0, \nonu \\
& \pa_{t_2}\bar {F}_1 + a_2(\pa^2_{x} \bar {F}_1 -2 b_1\bar b_1 \bar  {F} _1) + o_{-3} =0,
\label{e4}
\er
where
\br
g_{-1}&=&-\eta \psi e^ { {1\o 2}(\phi_{1} + \phi_{2}) },   \\
m_{-1}&=& \eta ( \chi + f_1 f_2 + \psi \chi f_1 f_2 + \chi f_2 g_2  + \psi \chi^2) e^{ -{1\o 2}( \phi_{1}+\phi_{2}) }, \nonu \\
n_{-1} &=& - \eta (g_2 + \psi f_1) e^{{1\o 2} \phi_1}, \nonu \\
o_{-1} &=& \eta f_2 (1+\psi \chi ) e^{-{1\o 2} \phi_1}
\er

Following the same argument as in the pure bosonic case, the vacuum configuration is obtained  from
\br T_0 = \exp (x  E^{(1)}) \exp \( (\a_2 E^{(2)} + {{\eta}} E^{(-1)}) t\)
\label{s221}
\er 
which leads to space-time dependence 
\br
\rho_i = \exp \(k_{i}x \) \exp  \( -(\a_{2} k_i^{2} + {{\eta}\o { k_i}} ) t\). 
\label{rhosuper}
\er
Following the soliton solutions for the Lund-Regge  model  obtained in \cite{silka} we have verified  solutions for eqns. (\ref{e4}) to be
 \br
 b_1 = \frac{\g_1 \rho_1^{-1} }{\tau_{0}},\qquad 
\bar b_1 = -\frac{\g_2 \rho_2}{\tau_{0}},  \qquad 
F_1 = -a_2\frac{\g_2 \rho_2}{\tau_{0}}, \qquad 
\bar F_1 = a_1\frac{\g_1 \rho_1^{-1} }{\tau_{0}}, \nonu 
\er
\br
\psi = {\frac{\rho_1}{\tau_0}}\( 1- {\frac{b\g_1 \rho_1^{-1} 
\rho_2}{2(\g_1-\g_2)(1+{\frac{\g_1}{\g_2}} \rho_1^{-1}  \rho_2)}}\), \qquad 
\chi = {\frac{\rho_2}{\tau_0}}\( 1- {\frac{b\g_2 \rho_1^{-1} 
\rho_2}{2(\g_1-\g_2)(1+{\frac{\g_1}{\g_2}} \rho_1^{-1}  \rho_2)}}\), \nonu \er
\br
g_1 = a_2{\frac{\g_1 \rho_1^{-1} 
\rho_2}{(\g_1-\g_2)\tau_0}}e^{-\frac{1}{2}\phi_1}, \qquad 
{f_1} = a_1{\frac{\g_1 \rho_1^{-1}  \rho_2}{(\g_1-\g_2)\tau_0}}e^{-\frac{1}{2}\phi_1}, \qquad 
g_2 = a_1{\frac{ \rho_1^{-1}  }{\tau_0}}e^{-\frac{1}{2}\phi_2},\qquad 
{f_2} = a_2{\frac{
\rho_2}{\tau_0}}e^{-\frac{1}{2}\phi_2},\nonu \er
\br
e^{\frac{1}{2}(\phi_1 + \phi_2)} = \frac{1+a_{3}\rho_{1}\rho_{2}}{\tau_{0}}, \qquad 
e^{\frac{1}{2}(\phi_1 - \phi_2)} =
\frac{1+\bar{a_{3}}\rho_{1}^{-1} \rho_{2}}{\tau_{0}},
\label{solrel}
\er 
 where $a_1,a_2$  and $b$ are Grassmaniann  and bosonic constants respectively, $\rho_i, i=1,2$ are given by (\ref{rhosuper}) and 
 \br
 a_3 = {{k_1}\o {k_2}}\Gamma_0 (1- b{{(k_1 +k_2)}\o {2k_1}}), \qquad \bar a_3 = \Gamma_0 (1+ b{{(k_1 -3k_2)}\o {2k_2}}),\nonu \\
  \Gamma = (1-a_1a_2)\Gamma_0,  \qquad  \Gamma_0 = {{k_1k_2}\o {(k_1 - k_2)^2}}, \qquad 
 \tau_0 = 1+\Gamma \rho_1^{-1} \rho_2.
 \er

 \section{General Case}
 
 We now  consider a mixed hierarchy  associated  to a general  affine Lie algebra $\hat \lie = \oplus \lie_i$, $[Q, \lie_i] = i \lie_i$ and  constant grade one semi-simple element $E$ such that $\hat \lie = { \cal M} \oplus {{ \cal K}}, \; [E, {\cal K}]=0$ with  the symmetric space structure,
 \br 
 \lb {{ \cal K}}, {{ \cal K}}\rb \subset {{ \cal K}}, \qquad \lb {{ \cal K}}, {{ \cal M}}\rb \subset {{ \cal M}}, \qquad \lb {{ \cal M}}, {{ \cal M}}\rb \subset {{ \cal K}}.  \label{90}
 \er
 with  equations of motion involving 
 time evolution with two indices, $t_{n,m}$ defined from the zero curvature representation
  \br
\lb  \pa_x + E + A_0, \pa_{t_{n,m}} +   D^{(n)} +  D^{(n-1)} + \cdots  D^{(0)} +D^{(-1)} +\cdots  D^{(-m+1)} +  D^{(-m)}\rb = 0. 
\label{8}
\er 
Eqn. (\ref{97})  leads to  
\br
&&\lb E,  D^{(n)}\rb = 0\label{91} \\
&&\lb E,  D^{(n-1)}\rb  + \lb A_0, D^{(n)}\rb + \pa_x D^{(n)} = 0\label{92} \\
&&\vdots \nonu \\
&&\lb E,  D^{(n-i)}\rb  + \lb A_0, D^{(n-i+1)}\rb + \pa_x D^{(n-i+1)} = 0\label{93}\\
&&\vdots \nonu \\
&&\lb E,  D^{(-1)}\rb + \lb A_0, D^{(0)}\rb + \pa_x D^{(0)} - \pa_{t_{n,m}} A_0 = 0 \label{94} \\
&&\lb E,  D^{(-2)}\rb  + \lb A_0, D^{(-1)}\rb + \pa_x D^{(-1)} = 0\label{95}\\
&&\vdots \nonu \\
&&\lb E,  D^{(-j-1)}\rb  + \lb A_0, D^{(-j)}\rb + \pa_x D^{(-j)} = 0\label{96} \\
&&\vdots \nonu \\
&& \lb A_0, D^{(-m)}\rb + \pa_x D^{(-m)} = 0\label{97}
\er
In order to solve eqns. (\ref{91})-(\ref{97}) we have to start from both ends, i.e. from (\ref{91}) towards (\ref{94}) , using the symmetric space structure (\ref{90}), 
we project each eqn. into $\cal K$ and $\cal M$ subspaces to obtain $ D^{(i)}_{\cal K}, i=1, \cdots n$ and $ D^{(i)}_{\cal M}, i=0 \cdots n$.   
On the other hand, starting from (\ref{97}) upwards, we find solution for $ D^{(-j)}_{\cal K}$ and $ D^{(-j)}_{\cal M}, j=1 \cdots m$ which is non-local in the fields in $A_0$.  For the particular case when $m=1$, we have seen that  there is a set of variables  within a group element $B$  that solves (\ref{97}) locally for $m=1$.

Inserting $ D^{(-1)}$ in (\ref{94}) and projecting in $\cal K$  we find $ D^{(0)}_{\cal K}$   which in turn determines the time evolution as the projection of (\ref{94}) in $\cal M$.  Following the same arguments given before, the space-time dependence of such generalized mixed model will be of the form
\br
\rho_i = \exp (k_i x) \exp \( (\a_i k_i^n + \eta k_i^{-m})t\)
\er

 \vskip 1cm \noindent

{\bf Acknowledgements} \\
\vskip .01cm \noindent
{  We thank CNPq for support.}


\bigskip

 \begin{thebibliography}{99}
 \bibitem{chodos} A. Chodos, \PRD{21}{1980}{2818}
 \bibitem{nissimov}H. Aratyn, J.F. Gomes, E. Nissimov, S. Pacheva and  A.H. Zimerman 
``Symmetry flows, conservation laws and dressing approach to the integrable models'', 
Proc. of the NATO Advanced Research Workshop on Integrable Hierarchies and Modern Physical 
Theories (NATO ARW - UIC 2000), Chicago (2000), nlin/0012042

\bibitem{modugno}H. Aratyn , J.F. Gomes and  A.H. Zimerman 
\JGP{46}{2003}{21}, hep-th/0107056
 
  \bibitem{konno}  K. Konno, W. Kameyama and H. Sanuki, \JPSJ{37}{1974}{171}
  \bibitem{chen}D-y. Chen, D-j. Zhang and S-f. Deng, \JPSJ{71}{2002}{658}

 \bibitem{leblond} H. Leblond, I.V. Melnikov and D. Mihalache, \PRA{78}{2008}{043802-1}
 \bibitem{laf} L. A. Ferreira, J.L. Miramontes and  J. Sanchez-Guillen,\JMP{38}{1997}{882},  hep-th/9606066
\bibitem{jpal} H. Aratyn, L. A. Ferreira, J.F. Gomes and A.H. Zimerman, \JPA{33}{L331}{2000},
 nlin/0007002
 \bibitem{wigner}I. Cabrera-Carnero, J.F. Gomes, E.P. Gueuvoghlanian, G.M. Sotkov, A.H. Zimerman, Proc. of the 7th International Wigner Symposium (Wigsym 7), College Park, Maryland, (2001), hep-th/0109117
 \bibitem{np} H. Aratyn, J.F. Gomes and A.H. Zimerman,\NPB{676}{2004}{537}, hep-th/0309099
 
 \bibitem{pla}J.F. Gomes, L.H. Ymai, A.H. Zimerman, \PLA{359}{2006}{630}, hep-th/0607107
 \bibitem{silka}H. Aratyn, J.F. Gomes, G.M. de Castro, M.B. Silka and  A.H. Zimerman 
\JPA{38}{2005}{9341}, hep-th/0508008





 \end{thebibliography}
\end{document}